\documentclass[twocolumn,prb]{revtex4}

\usepackage[utf8x]{inputenc}
\usepackage{psfrag}
\usepackage{color}
\usepackage{amsmath}
\usepackage{amssymb}
\usepackage{amsfonts}

\usepackage{ifpdf}
\ifpdf
\usepackage[pdftex]{color,graphicx}
\DeclareGraphicsExtensions{.pdf, .jpg, .tif, .png}
\else
\usepackage{color,graphicx}
\DeclareGraphicsExtensions{.eps, .jpg}
\fi

\usepackage{times}
\usepackage{lmodern}
\usepackage[T1]{fontenc}

\usepackage{graphicx}
\usepackage{dcolumn}

\newcommand{\diff}[2]{\frac{\partial #1}{\partial #2}}

\newcommand{\abinitio}{\emph{``ab-initio''}}

\begin{document}

\title{Lindemann Criterion and the Anomalous Melting 
Curve of Sodium}

\author{M. Martinez-Canales}
\email{wmbmacam@lg.ehu.es}
\author{A. Bergara}
\email{a.bergara@ehu.es}

\affiliation{Materia Kondentsatuaren Fisika Saila, Zientzia eta Teknologia Fakultatea, Euskal Herriko Unibertsitatea, 644 Postakutxatila, 48080 Bilbo, Basque Country, Spain }

\affiliation{Donostia International Physics Center (DIPC), Paseo de Manuel Lardizabal, 20018, Donostia, Basque Country, Spain}

\affiliation{Centro Mixto CSIC-UPV/EHU, 1072 Posta kutxatila, E-20080 Donostia, Basque Country, Spain}


\begin{abstract}
Recent reports of the melting curve of sodium at high pressure have shown that it has a very steep descent after a maximum of around 1000K at 31 GPa. This is not due to a phase transition. According to the Lindemann criterion, this behaviour should be apparent in the evolution of the Debye temperature with pressure. In this work, we have performed an \abinitio{} analysis of the behaviour of both the Debye temperature and the elastic constants up to 102 GPa, and find a clear trend at high pressure that should cause a noticeable effect on the melting curve.
\end{abstract}

\maketitle

\section{Introduction}

At low pressures, alkali metals adopt high-symmetry simple structures and exhibit nearly-free-electron (NFE) behaviour. Thus, physics textbooks have always described group I elements as simple metals. Recent work, however, has shown that, at high pressure, there is a significant deviation from that simple NFE behaviour\cite{li-pairing,nature-pairing,prb-alvaro-06}, due basically to electronic shell collapse \cite{sternheimer}. Apparently simple behaviour is not restricted to electronic properties or crystal structures. The melting curves of the alkalis reported by Bridgman \cite{bridgman26} in 1926 were also relatively simple, fitting well to the monotonically increasing Simon's equation: 

\begin{equation} \label{eq-simon}
\left( \frac{T_m}{T_0}\right) ^c = \frac{p+a}{a}
\end{equation}

	However, this simple law breaks at high pressure, when $T_m$ does not follow eq. (1). Further experiments performed on the heavy alkalis $Rb$ and $Cs$ \cite{bundy59, kennedy62} found maxima in the melting curves, at 5 GPa in $Rb$ and, in $Cs$, a double maximum near the transition from $Cs$ I to $Cs$ II. Sodium shows an even more blatant deviation from eq. (1). Recent diamond anvil cell (DAC) experiments performed by Gregoryanz et al. \cite{gregoryanz04} showed the striking behaviour of $Na$: $T_m$ had a maximum of $\sim 1000K$ around 31GPa, and it then steeply decreased to room temperature at 100 GPa. 

The usual ab-initio approach to melting point calculation is based on phase coexistence simulations via first-principles molecular dynamics (FPMD). However, this is complicated and computationally extremely expensive. Simpler models have been used successfully in the past with the goal of describing the melting curve of several materials \cite{melt-alfe,melt-ahuja}. However, these materials do not deviate significantly from the behaviour described by Simon's Law. Can a simpler calculation, based on semiempirical criteria, reproduce the $Na$ melting curve? This work is oriented towards answering this question, given that simple criterions such as the Lindemann criterion or the Born stability criterion need a very definite behaviour to satisfy the experimental melting curve of $Na$.

\section{Computational Method}

The melting temperature calculations presented here will be based on the Lindemann melting criterion\cite{lindemann}. This model, based on the harmonic approximation, predicts that melting will occur when the root mean square displacement reaches a certain value (generaly about 1/8th) of the mean interatomic distance. Numerically,

\begin{equation} 
T_m= C v^{2/3}\Theta_D^{\phantom{D}2} \label{eq-lind}
\end{equation}

Although reasonable at first sight, this model has several defects. The assumption of harmonical forces discards the bond breakage occurring in melting, and it does not take into account the liquid phase. On the contrary, both the liquid and solid phases are related to the melting temperature and its evolution with pressure via the Clausius-Clapeyron equation.  Nevertheless, given the success various forms of the criterion have had (see e.g. \cite{melt-ahuja}), it is certainly interesting to test it in sodium and its characteristic melting curve. 

It is not very wise to take the meaning of the constant in equation \ref{eq-lind} very seriously, so the Lindemann criterion will be used here as a single-parameter model. Being interested in using the least possible \textit{external knowledge}, the free constant $C$ can be calculated from a single point. Then, working out the evolution of $T_m$ requires the knowledge of $T_m$(0 GPa) and the evolution of $\Theta_D$. Another approach, more useful to compare equation 2 with experiment, would be a least squares fitting of C to the experiment. However, the authors are more interested in using the least amount of parametres than in checking how good the fit of equation \ref{eq-lind} is. 

The Debye temperature has been evaluated from the phonon spectra at different pressures in two different ways. Once the phonon dispersion is known, one could obtain $\Theta_D$ from the phonon density of states (DOS). Applying the Debye approximation to the harmonic crystal yields the well known DOS:
\begin{equation}\label{eq-debyedos}
g_D(\omega) = \begin{cases}
\displaystyle \frac{3}{2\pi ^2}\frac{\omega ^2}{c ^3} & \text{if $\omega < \omega_D = k_D c$} \\
\displaystyle 0 & \text{if $\omega > \omega_D$,}
\end{cases}
\end{equation}
where $k_D$ is chosen to obtain the correct number of states. This condition is fulfilled if $k_D^{\phantom{D}3}=6\pi^2N/V$, where $N$ represents the numer of atoms per unit cell and V is the volume of the unit cell. The average speed of sound $c$ is obtained by fitting a line to the low $\omega$ region in a $g(\omega)$ versus $\omega^2$ plot. The Debye temperature is then easily obtained from its definition, $k_B \Theta_D = \hbar c k_D$. This is correct, since the approximations made in the Debye model are negligible in the low $\omega,T$ region.  

It is also possible to evaluate $\Theta_D$ from the elastic constants. In a cubic crystal, the low $\xi$ frequencies in the $(\xi, 0 ,0)$ branch and the elastic constats are related in the following manner:
\begin{equation} \label{eq-x00}
C_{11} = \rho \frac{\omega_l^{\phantom{l}2}}{\xi^2} , \qquad
C_{44} = \rho \frac{\omega_t^{\phantom{t}2}}{\xi^2} 
\end{equation}
In a similar fashion for the $(\xi, \xi, 0)$ branch,

\begin{align}\label{eq-xx0}
C_{11} - C_{12} = 2 \rho \frac{\omega_{t1}^{\phantom{t1}2}}{\xi^2},& {}\qquad
C_{44} = \rho \frac{\omega_{t2}^{\phantom{t2}2}}{\xi^2} \nonumber \\
C_{11} + C_{12}+ 2C_{44} = {}& 2 \rho \frac{\omega_l^{\phantom{l}2}}{\xi^2}
\end{align}
where $\omega_{t1}$ refers to the transversal mode contained in the $z=0$ plane. Symmetry considerations make these the only independent elastic constants in cubic crystals. This allows to readily obtain the bulk and shear moduli,

\begin{align}\label{eq-moduli}
B = {}& \frac{1}{3} (C_{11}+2C_{12}) \nonumber \\
G = {}& \frac{1}{5} (3C_{44}+C_{11}-C_{12})
\end{align}

Both moduli determine the long wavelegnth speed of longitudinal and transversal waves in a solid, via the following equations:

\begin{equation}\label{eq-refv}
v_t =  \sqrt{\frac{G}{\rho}}, \qquad v_l =  \sqrt{\frac{B+\frac{4}{3}G}{\rho}}
\end{equation}

In the Debye model of specific heat, one characterizes the behaviour of the material with $\Theta_D$, related to the vibrational properties of a solid via an average sound velocity. The two can be related through the following relation,

\begin{equation}\label{eq-vdeb}
\Theta_D = \frac{h}{k_b}\sqrt[3]{ \frac{3}{4\pi} \left( \frac{N_A\rho}{M} \right) } v_m
\end{equation}
where the average sound velocity $v_m$ can be defined as follows:
\begin{equation}\label{eq-vm}
\frac{3}{v_m^{\phantom{m}3}}= \frac{2}{v_t^{\phantom{t}3}} + \frac{1}{v_l^{\phantom{l}3}}
\end{equation}

So obtaining the Debye temperature by any of these two methods only require knowledge of the phonon spectrum. This has been done fully \abinitio, using density-functional perturbation theory (DFPT\cite{dfpt-zein,dfpt-baroni,dfpt-gonze}) as implemented in the Quantum-ESPRESSO code \cite{dfpt,pwscf}. It is important to note that phonons have been calculated in a regular mesh of the Brillouin zone. Then the interatomic force constant matrix has been obtained by interpolating the results to a Fourier series. This method has been preferred because it then allows ready calculation of phonon DOS and frequencies for an arbitrary $\vec{q}$. Convergence has been checked comparing the frequencies for selected $\vec{q}$, where direct calculations were available.

\section{Melting Curve of Sodium}

To study sodium, an LDA\cite{sic} norm-conserving pseudopotential has been considered satisfactory. Convergence to the desired accuracy has been achievd with a 30 Ry cutoff. Similarly, the summations over the Brillouin zone have been performed in a, at worst, 18 18 18 Monkhorst-Pack\cite{monkhorst-pack} grid. In some cases, up to 22 22 22 grids have been used. Another important point is the phonon grid used to interpolate the interatomic force constant (IFC) matrix. With the goal of getting a quality IFC matrix,  a $\Gamma$-centered 10 10 10 phonon grid has been used for both bcc and fcc $Na$. The quality of the IFC matrix has been checked by comparing the frequencies obtained directly from DFTP calculations and the interpolated values at selected $\vec{q}$-points.

\begin{table}
  \centering
    \begin{tabular}{c c c c c c c}
\hline
\hline
  P (GPa)  &  $a (a_0)$   &  $C_{11}$& $C_{12}$  & $C_{44}$  & \multicolumn{2}{c}{ $\Theta_D$ (K)  }\\   
\hline
 0.00& 7.738& 9.79&  7.47&  6.72&195.0& 147.7 \\
10.53& 6.60 & 43.9&  36.9& 14.1 &284.5& 229.0 \\
20.93& 6.20 & 73.8&  62.9& 17.0 &308.7& 253.2 \\
26.73& 6.05 & 88.8&  76.8& 17.9 &316.2& 260.5 \\
29.91& 5.98 & 97.2&  84.6& 18.2 &318.6& 262.8 \\
43.11& 5.75 &131.2& 115.3& 18.7 &323.5& 260.1 \\
52.08& 5.63 &153.2& 135.0& 18.4 &323.5& 255.9 \\
63.50& 5.50 &182.7& 162.4& 17.6 &319.4& 249.8 \\
\hline
63.50& 6.94 &179.9& 170.3& 17.8 &297.0& 245.9 \\
73.12& 6.82 &204.8& 194.7& 16.3 &285.1& 234.1 \\
84.91& 6.70 &233.5& 222.1& 14.1 &269.7& 217.5 \\
93.83& 6.62 &255.1& 242.4& 12.1 &256.2& 203.8 \\
102.4& 6.55 &275.7& 263.3&  9.9 &236.3& 174.9 \\
\hline
\hline
    \end{tabular}
  \caption{Evolution with pressure of the elastic constants (in GPa) and $\Theta_D$ of $Na$. The $\Theta_D$ on the left has been calculated via the elastic constants, while the rightmost one is calculated from the phonon DOS.\label{tab:na-const}}
\end{table}

    With the approximations and numerical parameters fixed as mentioned above, the phonon spectrum of $Na$ has been calculated from ambient pressure to 103 GPa, on the onset of the $fcc$ to $cI16$ phase transition. The phases probed have been $bcc$ (from 0 to 63 GPa) and $fcc$ (63 to 103 GPa). Table \ref{tab:na-const} shows the calculated evolution of the elastic constants with pressure.

\begin{figure}[hbt]
  \centering
    \includegraphics[width=0.46\textwidth]{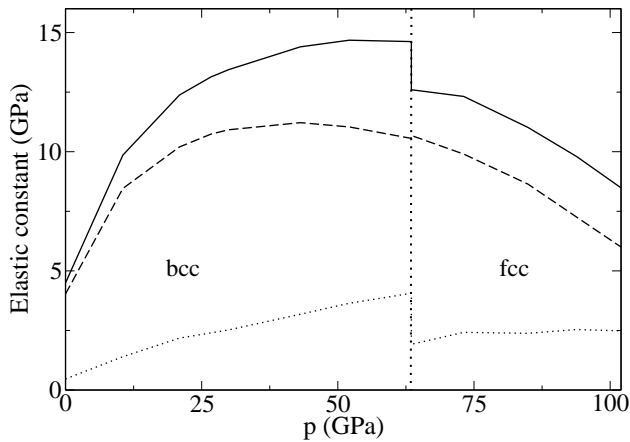}
  \caption{Evolution with pressure of the $Na$ shear modulus (solid line), together with the contribution of $C_{44}$ (dashed line) and $C_{11}-C_{12}$ (dotted line). \label{fig-nashear}}
\end{figure}

    The first thing one notices is that $C_{44}$ reaches a maximum at around 43 GPa and, after the phase transition, it decreases dramatically. Although $C'=\frac{1}{2}(C_{11}-C_{12})$ increases, its growth can not compensate the decrease of $C_{44}$, which causes the decay of the shear modulus. As a consequence of this, $\Theta_D$ has a maximum in this pressure range, which, according to equation \ref{eq-lind}, should also result in a maximum in the melting curve of sodium.

\begin{figure}[hbt]
\centering
  \includegraphics[width=0.48\textwidth]{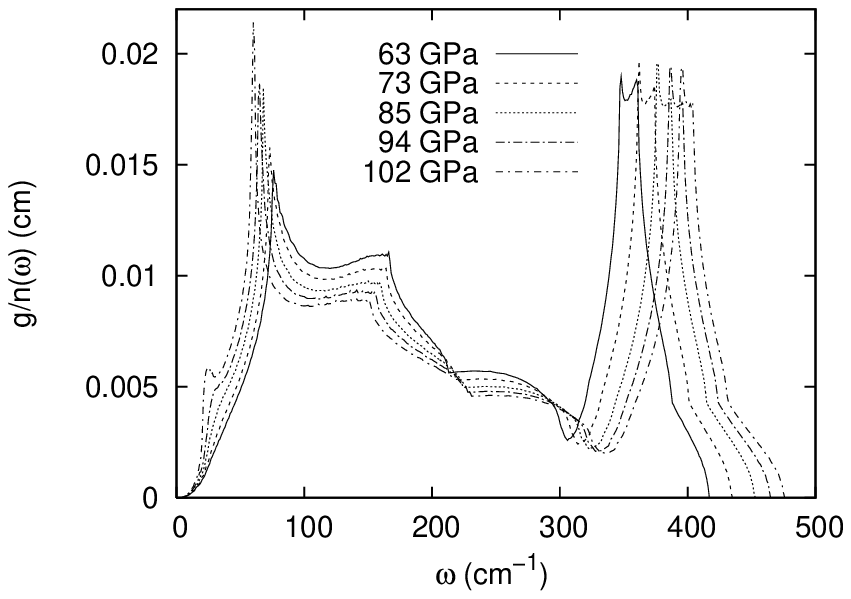}
  \includegraphics[width=0.48\textwidth]{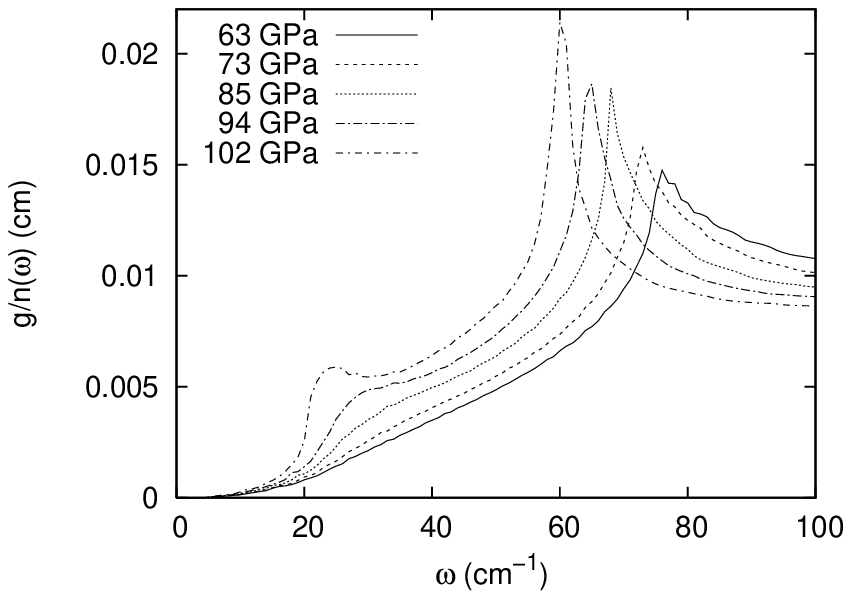}
  \caption{Evolution with pressure of the phonon DOS in $fcc$ $Na$. A detail of the low $\omega$ region is also depicted. \label{fig-naphdos}}
\end{figure}

We now turn our attention to the evolution of the phonon DOS, represented in figure \ref{fig-naphdos}. There, one can appreciate that the high-frequency region of the spectrum, consisting basically on longitudinal modes, increases with pressure. On the other hand, the low frequency van-Hove singularity corresponding to transversal modes decreases monotonically. From equations \ref{eq-x00}, \ref{eq-xx0} and \ref{eq-moduli} it is possible to appreciate that the shear modulus of $fcc$ $Na$ is collapsing, a  similar behvaviour to that described by Kechin \cite{kechin-shear}. What the Lindemann criterion suggests is that this has very definite consequences on the evolution of $T_m$ with pressure. From equation \ref{eq-vm}, on the limit $v_l >> v_t$, the average sound velocity $v_m$ will tend to $v_t$. Substituting $v_t$ into equation \ref{eq-vdeb} results in $\Theta_D \propto \sqrt{G} \rho^{-1/6}$. The resulting decrease of $\Theta_D$, together with the theromdynamic stability condition of $\diff{V}{P} < 0$ means, according to equation \ref{eq-lind} that the melting temperature will decrease.

\begin{figure}[hbt]
  \centering
  \includegraphics[width=0.47\textwidth]{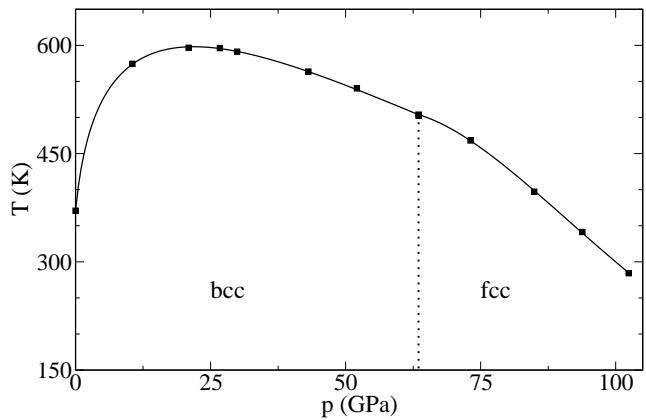}
  \caption{Melting curve of $Na$. The shown line is a least squares fit to the Kechin equation \cite{kechin-melting}\label{fig-namelt}}
\end{figure}

With these considerations in mind, the behaviour of the calculated melting curve is not surprising. The melting curve, calculated from the elastic constants, is shown in figure \ref{fig-namelt}. The first thing to notice is how a simple criterion is able to predict qualitatively the behaviour of the melting curve of $Na$. It also predicts that, at 102 GPa, $Na$ should be a liquid at ambient temperature. However, this method also has its shortcomings. The most noticeable is that, even though it predicts that the melting curve will have a maximum, it severely underestimates the maximum at 600K. Even if one takes into account the large error margins presented by Gregoryanz \textit{et al.}\cite{gregoryanz04} and the known shortcomings of LDA, this is still a large difference. In any case, it is encouraging that the theoretical maximum is located at 25 GPa, very close to the pressure of the experimental maximum, 31 GPa. This difference can even be caused by the slow convergence of phonon frequencies for small $\vec{q}$. 

It should be noted, however, that the free parameter in the Lindemann model has been determined from a single point. Another option would have been to use least-squares fitting to the experimental data, or even to the maximum of the curve, which would have resulted in better fits. This has not been done here because the authors were interested in using the least amount of external information.

\section*{ Conclusion}

   In this work we present an \abinitio{} analysis of the evolution of the phonon DOS and the elastic constants of $Na$ with pressure. It has been shown that for pressures higher than 40 GPa the Debye temperature $\Theta_D$ and the shear modulus start decreasing. This decrease is even steeper after the $bcc$ to $fcc$ transition. Additionally, an estimate of the melting curve of $Na$ has been made, with the only knowledge of the melting temperature  of sodium at ambient pressure.

   Another interesting conclusion one can draw from the results is that, although naïve, the Lindemann criterion supplemented with \emph{ab-initio} phonon DOS correctly predicts the existence of a maximum in the melting curve of sodium, at a pressure remarkably close to the experimental figure. The predictions based on the criterion fail to predict a high enough melting temperature, although this could be aggravated by the slow convergence of phonons for small $\vec{q}$. One might be worried that it cannot predict the flat slope of the melting curve on the beginning of the fcc-liquid part shown in \cite{gregoryanz04}. In any case, there were no experimental points reported in that region. Furthermore, more elaborate calculations presented in \cite{melt-iniguez} also support the idea that $\diff{T_m}{p}<0$ in that region.

   Although using the $\Theta_D$ from either the phonon DOS or the elastic constants does not result in qualitatively different results, it seems clear from Fig. \ref{fig-naphdos} that an approach based on $\langle \omega \rangle$ will probably fail in sodium, as it increases with pressure.

   Given the qualitative agreement found for $Na$, one might be tempted to do some qualitative predictions for lithium, for which technical difficulties have slowed experimental advance. Preliminary calculations\cite{masmelting} show no collapse of the shear modulus, neither a decrease of the Debye temperature up to 35 GPa. Taking into account the underestimate of the melting temperature for sodium and that at low pressures $T_m$ should have a subtle positive slope \cite{bridgman26}, it seems most likely that $T_m$ will rise slowly until the bcc to fcc phase transition, where a Rb-like maximum might appear. 

\bigskip
The reesearch herein presented was performed under project BFM2003-04428, funded by th Spanish Ministry of Education and Science. One of us, M.M.C. wants to thank the Spanish Ministry of Education and Science for economic support and grant BES-2005-8057. Finally, the authors are also thankful to SGI IZO SGIker UPV EHU for the allocation of computational resources.

\bibliography{arxiv}

\end{document}